\documentclass[aps,prl,twocolumn,superscriptaddress,groupedaddress]{revtex4}  

\usepackage[utf8]{inputenc}
\usepackage{graphicx}  
\usepackage{dcolumn}   
\usepackage{natbib}
\usepackage{capt-of}
\usepackage {booktabs}
\usepackage[caption=false]{subfig}
\usepackage{bm}        
\usepackage{amssymb}   
\usepackage{amsmath}
\usepackage{hyperref}
\hypersetup{
    colorlinks=true,
    linkcolor=blue,
    filecolor=blue,
    citecolor=blue,
    urlcolor=blue,
}
\usepackage{color}
\usepackage{comment}
\begin{document}
\title{Experimental generation of extreme electron beams for advanced accelerator applications}
\author{C. Emma}
\affiliation{SLAC National Accelerator Laboratory, Menlo Park, California 94025, USA}

\author{N. Majernik}
\affiliation{SLAC National Accelerator Laboratory, Menlo Park, California 94025, USA}

\author{K. K. Swanson}
\affiliation{SLAC National Accelerator Laboratory, Menlo Park, California 94025, USA}

\author{R. Ariniello}
\affiliation{SLAC National Accelerator Laboratory, Menlo Park, California 94025, USA}

\author{S. Gessner}
\affiliation{SLAC National Accelerator Laboratory, Menlo Park, California 94025, USA}

\author{R. Hessami}
\affiliation{SLAC National Accelerator Laboratory, Menlo Park, California 94025, USA}

\author{M. J.~Hogan} 
\affiliation{SLAC National Accelerator Laboratory, Menlo Park, California 94025, USA}

\author{A. Knetsch} 
\affiliation{SLAC National Accelerator Laboratory, Menlo Park, California 94025, USA}

\author{K. A. Larsen}
\affiliation{SLAC National Accelerator Laboratory, Menlo Park, California 94025, USA}

\author{A. Marinelli}
\affiliation{SLAC National Accelerator Laboratory, Menlo Park, California 94025, USA}

\author{B. O'Shea}
\affiliation{SLAC National Accelerator Laboratory, Menlo Park, California 94025, USA}

\author{S. Perez}
\affiliation{SLAC National Accelerator Laboratory, Menlo Park, California 94025, USA}

\author{I. Rajkovic}
\affiliation{SLAC National Accelerator Laboratory, Menlo Park, California 94025, USA}

\author{R. Robles}
\affiliation{SLAC National Accelerator Laboratory, Menlo Park, California 94025, USA}

\author{D. Storey}
\affiliation{SLAC National Accelerator Laboratory, Menlo Park, California 94025, USA}

\author{G. Yocky}
\affiliation{SLAC National Accelerator Laboratory, Menlo Park, California 94025, USA}

\begin{abstract}
In this Letter we report on the experimental generation of high energy (10 GeV), ultra-short (fs-duration), ultra-high current ($\sim$ 0.1 MA), petawatt peak power electron beams in a particle accelerator. These extreme beams enable the exploration of a new frontier of high intensity beam-light and beam-matter interactions broadly relevant across fields ranging from laboratory astrophysics to strong field quantum electrodynamics and ultra-fast quantum chemistry. We demonstrate our ability to generate and control the properties of these electron beams by means of a laser-electron beam shaping technique. This experimental demonstration opens the door to on-the-fly customization of extreme beam current profiles for desired experiments and is poised to benefit a broad swathe of cross-cutting applications of relativistic electron beams. 

\end{abstract}

\maketitle

Particle accelerators are among the most complex and versatile tools in existence for probing the natural world \cite{Sessler2014}. Progress in the study and understanding of the fundamental forces and interactions of nature has been predicated on continuing improvements in accelerator beam brightness, intensity, and customization. Next generation experiments spanning a wide variety of disciplines will require beams with unprecedented properties: 10s - 100s of nm bunch length, MA-scale peak current, and peak electric fields exceeding 1  V/$\AA$.  These experiments will push the boundaries of our understanding in areas ranging from high energy physics \cite{Yakimenko2019}, strong field quantum electrodynamics \cite{Bulanov2022,Matheron2022}, laboratory astrophysics \cite{Benedetti2018}, ultrafast photo-chemistry and material science \cite{Cesar2022} to highlight a few examples. Furthermore, ultra-short, ultra-high current beams are envisioned as drivers for future light sources \cite{Xu2021,Emma2021}, advanced particle accelerators \cite{Rosenzweig2011} and laserless gamma-gamma colliders \cite{Blankenbecler1988,ShortBunchParadigm2021}. The success of all these efforts depends on the generation of extreme beams with tailored current profiles and orders of magnitude higher beam densities beyond what is achievable at state-of-the-art accelerator facilities to date.  To this end, the controlled generation of ultra-high current beams with an eye towards reaching solid density bunches has been recognized as one of the `grand challenges' of particle accelerator and beam physics \cite{abp_workshop_2023}.

In this Letter we report on the experimental generation of high energy (10 GeV), ultra-short (fs-duration), extremely high peak current (0.1 MA), petawatt (PW) peak power electron beams in a particle accelerator. The generation of these bunches represents an $\sim$5x increase in previously achieved maximum field strength using state-of-the-art relativistic particle beams \cite{Corde2016}, enabling the next generation of experiments exploring high-intensity beam interactions with light and matter. We generate such high peak current beams using the controlled shaping of the electron energy profile with an external, spectrally-modulated, ps-duration infrared (IR) laser pulse. This approach exploits the combination of laser shaping and the beam's self-interaction to generate a fs-current spike containing hundreds of pC charge \cite{Cesar2021,Li2024}. Our experiment leverages recent developments in electron beam shaping developed for femtosecond and attosecond pulse control at x-ray free-electron lasers \cite{Ackermann2007,Emma2010,Ishikawa2012,Allaria2013,Kang2017,Prat2020,Liu2021,Wang2021}, and improves the peak current by one order of magnitude compared to previous experimental results \cite{Duris2019,Duris2021,Cesar2021,Li2024,Huang2017,Zhang2020,Yan2024}.   

We characterize the charge, peak current, and temporal duration of these high current beams and show that we can utilize them to trigger and control the onset of beam-induced ionization of gas targets broadly relevant to advanced accelerator applications. We measure the impact of the beam-ionized plasma on the electron energy distribution and correlate it to the energy deposited in the plasma via measurements of the plasma afterglow \cite{Oz2004,Scherkl2022,Boulton2022,Zhang2024arxiv} and the gamma ray yield from betatron radiation emitted by the beam as it transits through the plasma \cite{SanMiguelClaveria2019}. Finally, we demonstrate our ability to control the amplitude and temporal location of the current spike in the beam by varying the shaping laser parameters.

\begin{figure*}[htbp]
    \centering
    \includegraphics[width=\linewidth]{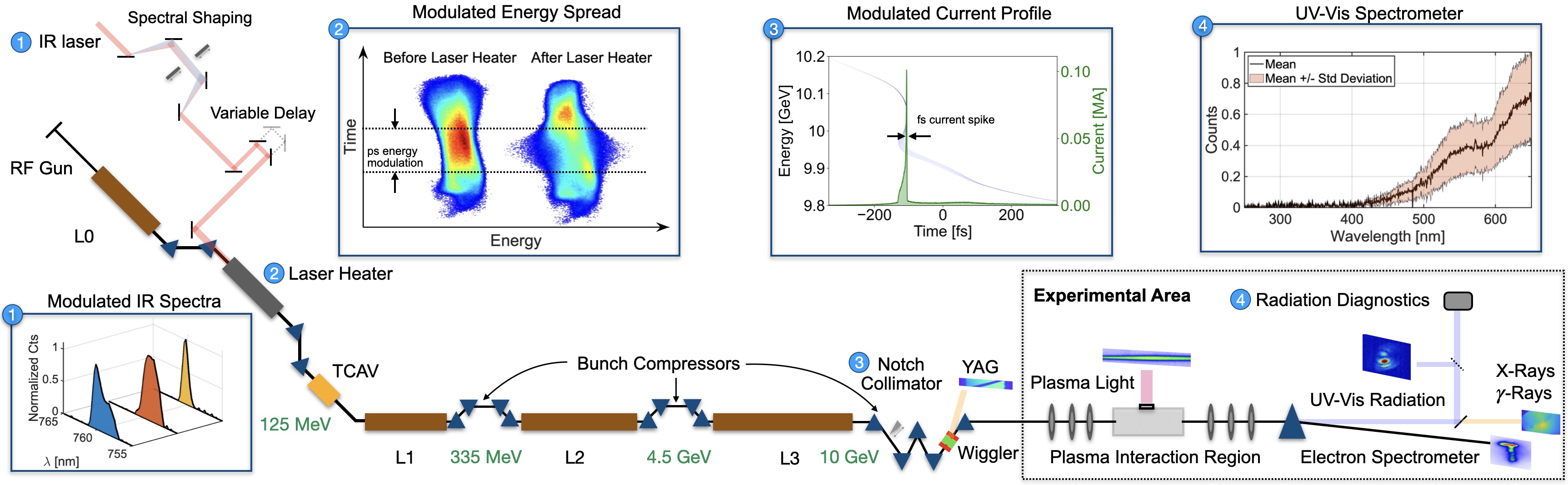}
    \caption{Schematic of the ultra-high current, extreme beam generation experiment at FACET-II. (1-2) A spectrally-modulated infrared laser imparts an energy modulation on the electron beam. (3) Space charge and successive bunch compressors transform the energy modulation into a current modulation at the entrance of the experimental area. (4) The charge, temporal duration and peak current of the resulting fs-duration, 0.1 MA current spike is measured via notch collimator scans, coherent radiation spectroscopy, and through beam-induced ionization of a meter-scale gas target.}
    \label{fig:Fig1Schematic}
\end{figure*}

The experiments presented in this work were performed at the FACET-II National User Facility at SLAC National Accelerator Laboratory \cite{YakimenkoFACETII2019,Storey2024}. A schematic of the experimental setup is shown in Fig. \ref{fig:Fig1Schematic}. The FACET-II facility provides beams with variable charge (nominally 0.5 - 3.0 nC) from a 125-MeV high brightness photoinjector and accelerates them to 10 GeV energy in a series of normal conducting S-band accelerating cavities interspersed with three bunch compressors. The critical hardware component which enables the generation of fs-duration current spikes described in this work is the laser heater, which allows controlled modulation of the electron beam energy spread at the exit of the photoinjector. The laser heater is composed of a 9 period undulator located in the middle of a 4-dipole chicane in which a ps-duration pulse from a Ti:Sapphire laser system is overlapped with the electron bunch to modulate the electron energy spread. Traditionally, this laser heating is used in high brightness linacs to damp the microbunching instability and suppress the formation of sub-structure in the electron beam current profile that can lead to a degradation of the beam quality \cite{Heifets2002,Saldin2002,Saldin2004,Huang2004,Akre2008,Huang2010,Ratner2015}. 

\begin{figure}[htbp]
    \centering
    \includegraphics[width=\linewidth]{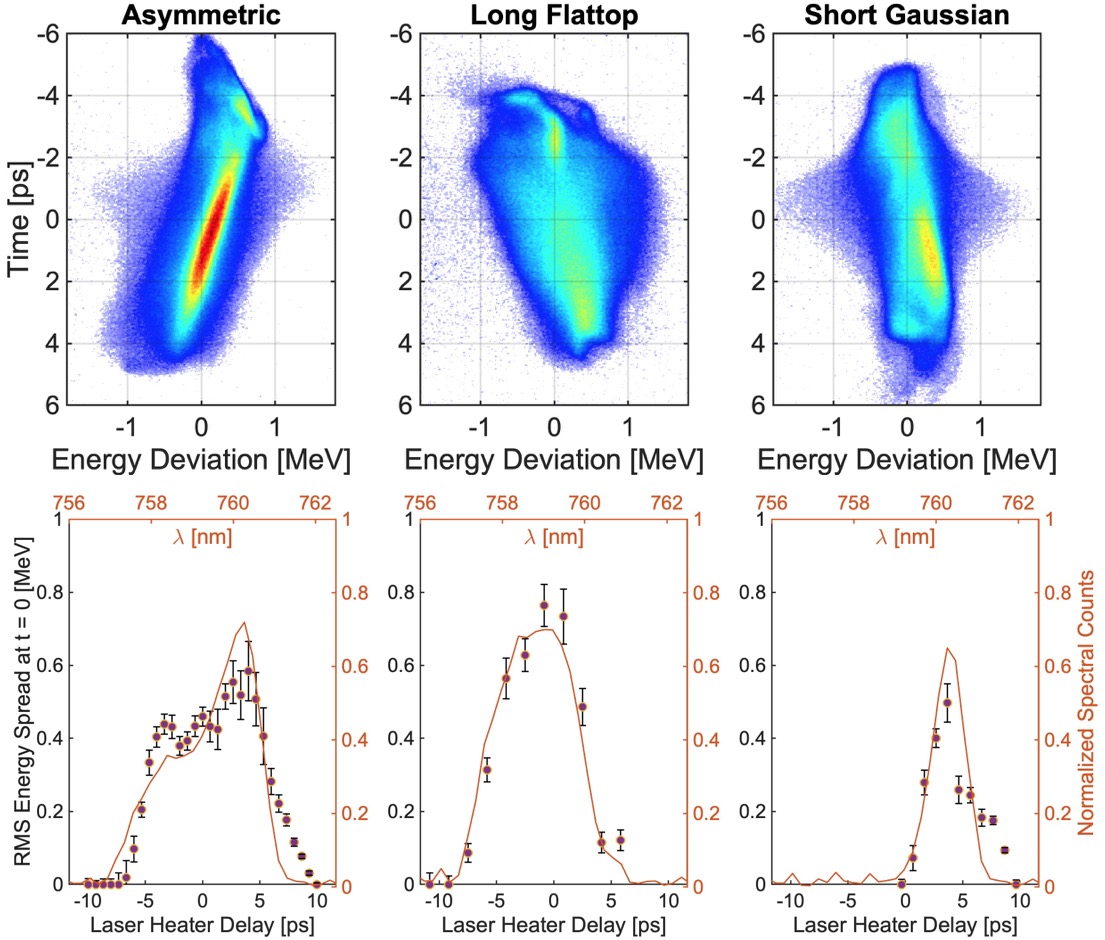}
    \caption{(Top row) Example longitudinal phase spaces at the exit of the laser heater and corresponding increase in the slice energy spread (bottom row) for different modulated laser spectra. }
    \label{fig:Fig2HeaterModulation}
\end{figure}

Controlling the parameters of the laser pulse in the laser heater can  modulate the electron energy profile and shape the electron current distribution \cite{Lechner2014,Roussel2015,Marinelli2016,Grattoni2017}. The laser heater system at FACET-II allows independent control of the spectro-temporal profile of the IR pulse, providing broad access to pulse shaping and current profile manipulation. In our experiment, spectral masks \cite{Heritage1985} were employed to generate spectrally and temporally modulated heater profiles resulting in modulated energy spread profiles at the laser heater exit. Example modulated energy spread profiles are shown in Fig. \ref{fig:Fig2HeaterModulation}. This figure depicts the electron beam longitudinal phase space (LPS) as measured at the exit of the laser heater using a vertically streaking transverse deflecting cavity and a horizontally deflecting dipole. The resulting LPS images along with the calculated RMS energy spread and corresponding IR spectra are shown for three different spectrally shaped IR pulses: an asymmetric profile, a long flattop profile and a short Gaussian. 

Different longitudinal modulations can generate desired current profiles at the experimental area. As described in \cite{Cesar2021}, in a simplified model considering a laser heater modulation followed by a single bunch compressor, the current profile at the exit of the bunch compressor is proportional to the second time derivative of the laser power. In the case of a more complex setup with multiple bunch compressors such as the FACET-II linac, particle tracking simulations \cite{myNote} are used to elucidate the beam dynamics through acceleration and transport which lead to the generation of isolated high current spikes as shown in Fig. \ref{fig:Fig1Schematic}. In our experiment, the modulated electron beam is transported from the exit of the laser heater through three additional accelerating sections and bunch compression stages. The final bunch compressor is equipped with a notch collimator which enables the selective removal of sub-sections of the electron beam to further shape and/or diagnose the bunch profile before it enters the experimental area. We utilize a thin (300 $\mu$m wide) notch scanning its horizontal position across the electron beam in a region inside the magnetic compressor where the beam is dispersed horizontally according to energy. This allows us to remove a fraction of the beam charge corresponding to a particular energy slice while leaving the remainder of the bunch undisturbed. In combination with the beam having an energy chirp at the collimator location, the transverse and temporal coordinates are correlated. Thus, removing particles within a certain range of horizontal positions is equivalent to removing particles within a certain range in both time and energy. Downstream of the notch collimator we measure the energy spectrum of the dispersed electron beam using a non-destructive energy spectrometer consisting of a half-period vertical wiggler which produces a streak of X-rays with an intensity profile that is correlated with the dispersed beam profile. The X-rays are intercepted by a scintillating yttrium-aluminum-garnet crystal and imaged by a charge-coupled device (CCD) camera with the resulting horizontal profile of the X-ray streak interpreted as the energy spectrum of the beam \cite{Seeman1986}.

\begin{figure}[t]
    \centering
    \includegraphics[width=\linewidth]{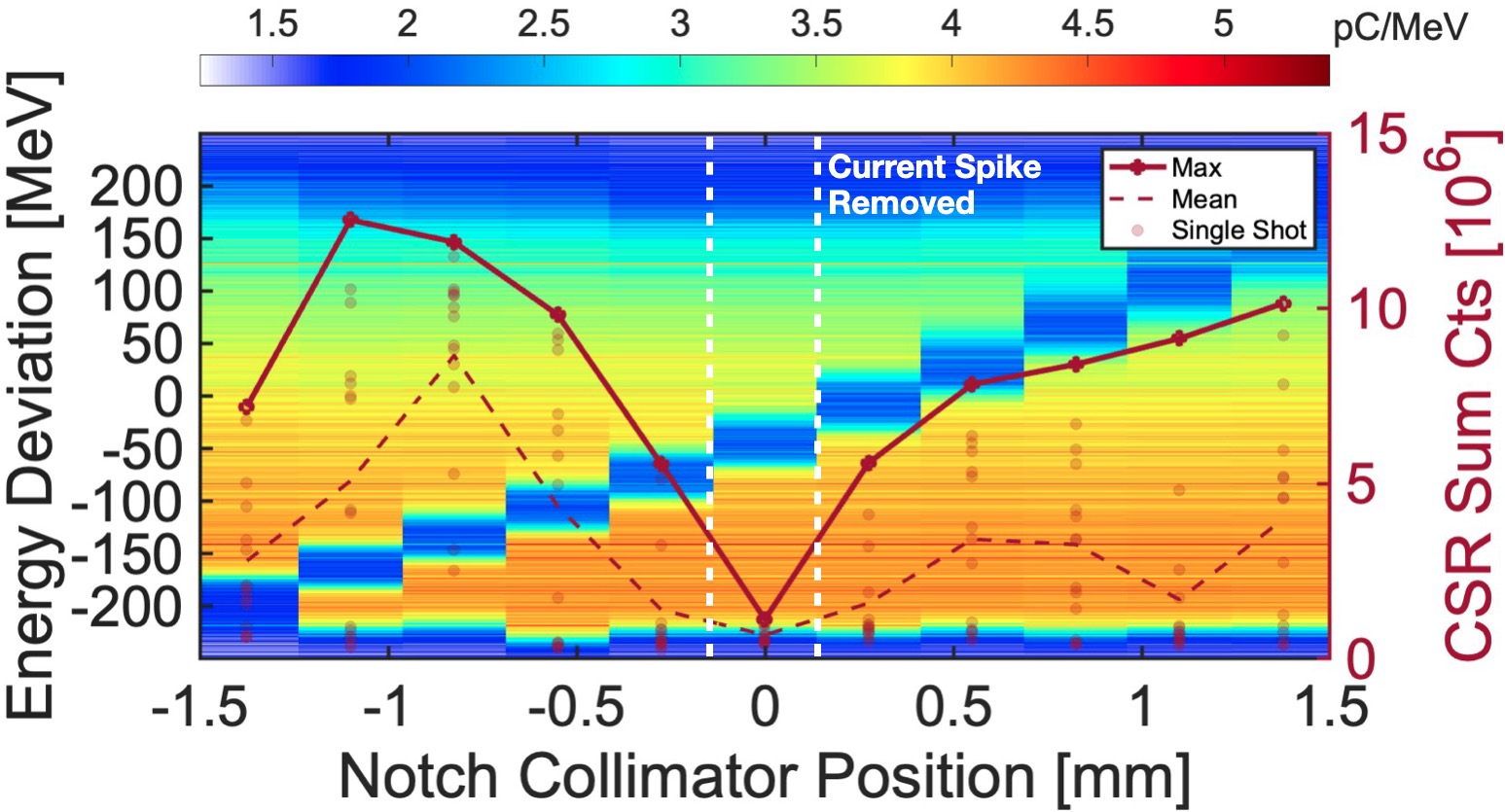}
    \caption{Notch collimator scan showing the average electron energy spectrum at each notch position and changes to the total amount of coherent radiation emitted by the beam in the experimental area. The false color plot represents the energy spectrum of the electron bunch, while the red dots and solid (dashed) lines are the single shot and maximum (mean) integrated CSR signal. The reduction in CSR emission at 0 mm reveals a short current spike highly localized in energy and time is present in the beam with $\sim$ 200 pC charge and is removed at a specific notch position.}
    \label{fig:Fig3NotchScans}
\end{figure}

During the course of the notch collimator scan we concurrently measure changes in the total energy radiated by the electron beam by imaging the coherent synchrotron radiation (CSR) emitted as the beam passes through the final dipole magnet in the experimental area on a CCD camera. An example notch scan measurement with a 1.6 nC charge beam and a short Gaussian IR laser modulation imparted at the laser heater is shown in Fig \ref{fig:Fig3NotchScans}. The figure is a waterfall plot showing the average energy spectrum (represented in false color) in each column as the notch is scanned across the beam. The notch removes $\sim$200 pC of charge from the electron beam at each position, and it is evident from the figure that the CSR signal (red line) is suppressed when the notch collimator position is near 0 mm and is intercepting an isolated current spike present in the beam. Other notch positions are characterized by frequent shots with large increases ($>10^6$) in the camera sum counts, where the notch is not intercepting the isolated current spike in the beam. Shot-to-shot variations in the CSR signal are the result of voltage and phase fluctuations of the RF accelerating structures in the FACET-II linac causing changes in the beam compression (see Ref \cite{TDR_FACET_II} for details).

\begin{figure}[t]
    \centering
    \includegraphics[width=\linewidth]{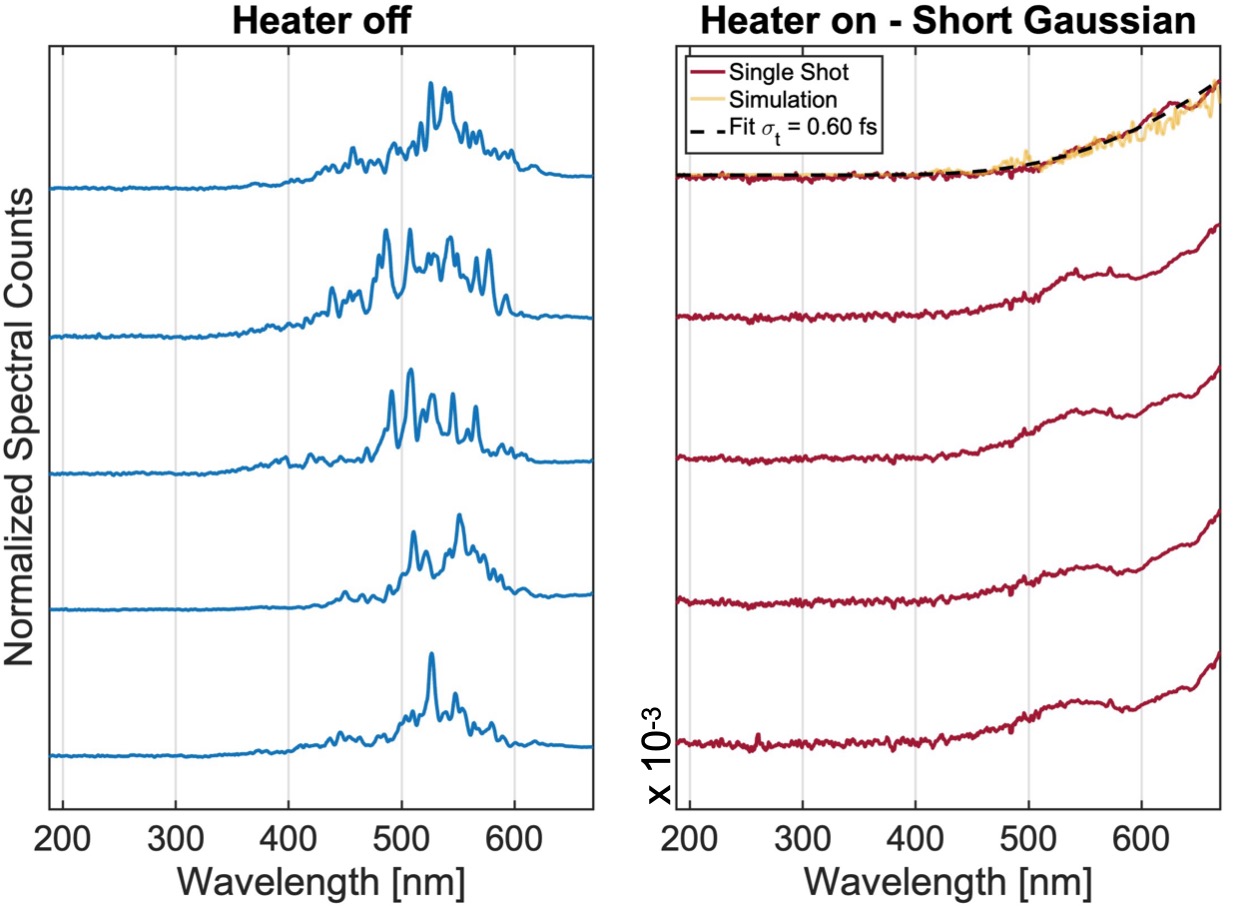}
    \caption{Representative shots of the UV-Visible coherent radiation spectra emitted by the beam with the laser heater off or on in the short Gaussian configuration (see Fig. \ref{fig:Fig2HeaterModulation}). The heater-off spectra are  characterized by stochastic emission seeded by shot noise in the electron beam and amplified by the microbunching instability. The heater-on spectra exhibit a smooth exponential decay indicative of the presence of a single high current spike of fs-duration in the bunch.}
    \label{fig:Fig4UVVisSpectra}
\end{figure}

To estimate the temporal duration of the isolated current spike in the beam, we measured the CSR radiation spectrum of the compressed bunch with the laser heater on and off. Resulting example spectra are shown in Fig. \ref{fig:Fig4UVVisSpectra}. The spectra with the heater on are multiplied by $10^{-3}$ to normalize for the increased signal level. The heater-off shots display a characteristic variation in the spectra resulting from stochastic changes in the bunching spectrum due to shot noise variation in the electron beam \cite{Lumpkin2002, Loos2008}. Figure \ref{fig:Fig4UVVisSpectra} shows the qualitative transition between shot-noise driven microbunching (large fluctuations and multiple spikes) to a single, laser heater-seeded current spike (smaller fluctuations with exponentially decaying spectra). For a temporally Gaussian current spike of duration $\sigma_t$, the coherent radiation spectrum of frequency $\omega$ scales as $I(\omega) \propto |b(\omega)|^2$ where $b(\omega) = \exp(-\omega^2 \sigma_t^2/2)$ is the bunching factor \cite{Gover2019}. The figure shows an exponentially decaying fit to the normalized spectra with $\sigma_t$ = 0.6 fs. A more detailed analysis based on particle tracking simulations \cite{myNote} reveals the profile of the ultra-short current spike is typically asymmetric, characterized by a slower rise time followed by a sharp drop and a full width at half maximum pulse duration $\Delta t = 3.2 $ fs. As shown in Fig. \ref{fig:Fig4UVVisSpectra}, the CSR spectra calculated using the bunching factor from simulated beam distributions are in good agreement with measured data.

We study the interaction of these high current beams with a 4 m-long static fill of He gas at density $n $ =  $1.65\times 10^{17}$cm$^{-3}$. Using the ADK model and particle-in-cell simulations \cite{Ammosov1986,Li2021}, we determine the magnitude of the electric field of the bunch required to ionize the He gas and, consequently, obtain an estimate of the peak current and transverse spot size of the beam to compare with our measurements. Simulations reveal that a peak current approaching 0.1 MA is required to ionize the He gas atoms from a transversely Gaussian electron with a focused spot size of $\sigma_{r}$ = 20 $\mu$m. These inferred parameters are consistent with measurements of the pulse duration obtained from CSR spectra and charge in the spike (as described above) as well as transverse spot size measurements of the beam in our experiments \cite{myNote}.

\begin{figure}[t]
    \centering
    \includegraphics[width=\linewidth]{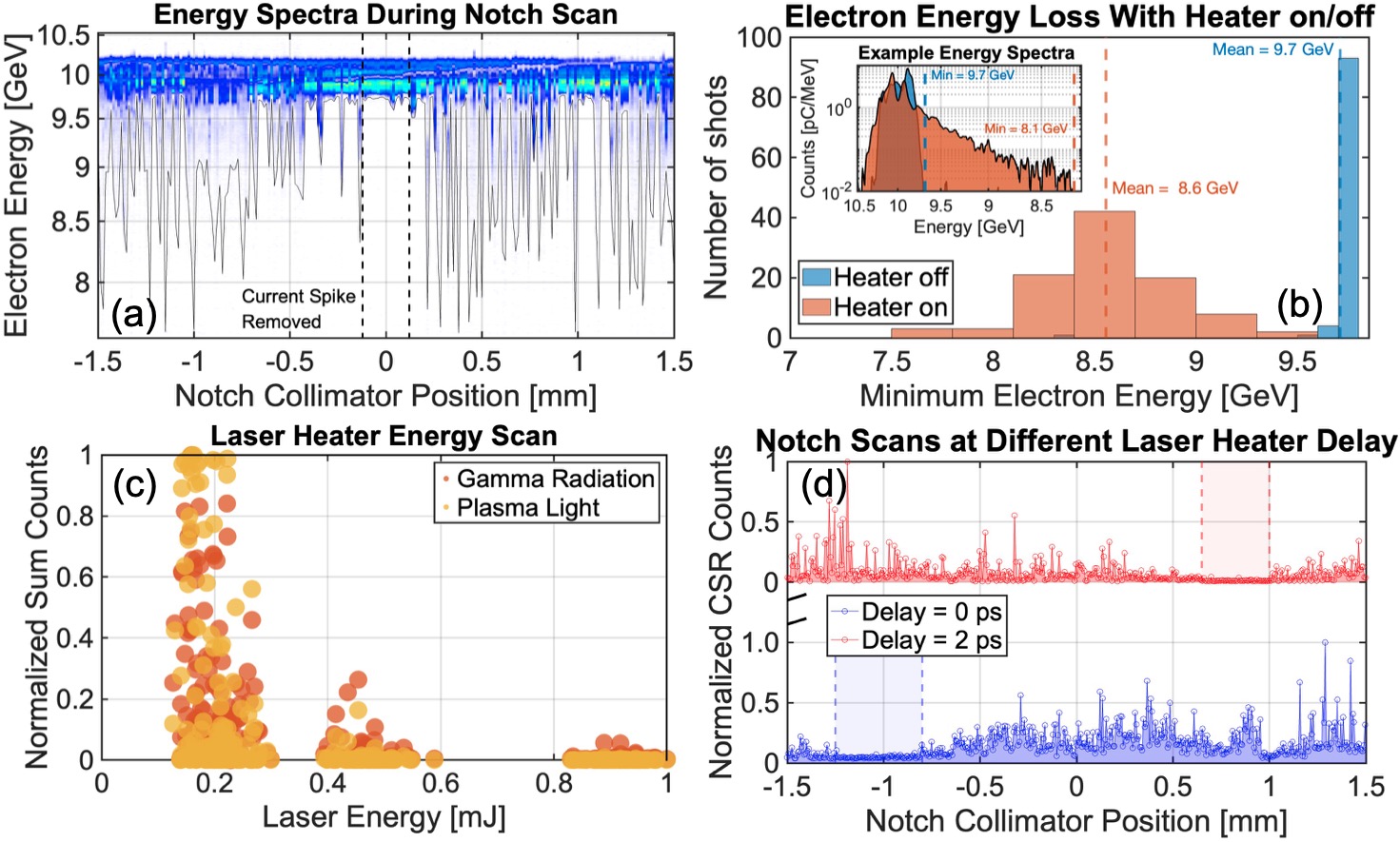}
    \caption{(a) - (b) Electron energy spectra with the beam incident on a static fill of He gas during a notch collimator scan with laser heater on/off for sequential shots. GeV-level energy loss due to beam-induced ionization of He is evident during heater-on shots and during the collimator scan when the notch is not removing the high peak current spike in the beam. (c) Plasma light and gamma ray emission during a laser heater energy scan showing control over the amount of ionization. (d) Notch scan at different laser heater delay stage positions demonstrating that the location of the spike in the beam can be controlled by the laser heater delay.}
    \label{fig:Fig5}
\end{figure}

The ionization of the gas by the electron beam causes the beam electrons to lose energy, transferring that energy to the He plasma. In Fig. \ref{fig:Fig5} (a) - (b) we show two separate datasets of the beam energy spectra, after interacting with the He static fill, measured with a dispersive electron energy spectrometer. Fig. \ref{fig:Fig5} (a) displays energy spectra taken during a notch collimator scan and exhibits a complementary trend to that visible in Fig. \ref{fig:Fig3NotchScans}, a majority of shots with some beam electrons experiencing large (GeV-level) energy loss and a number of shots that show negligible energy loss when the notch is positioned near 0 mm in such a way as to remove the isolated current spike from the beam. This confirms our ability to suppress the ionization of the He gas by intentionally removing the induced current spike and allows us to determine the position of the spike within the beam with high accuracy.  

Fig. \ref{fig:Fig5} (b) shows an example energy spectrum and the minimum energy of electrons in a dataset during which the laser heater is turned on/off on alternating shots by using a fast (3.5 ms opening time) shutter. The histogram illustrates the distribution of minimum electron energies in 198 consecutive shots. The distribution shows that beam electrons experience negligible energy loss with the heater off and a large  (GeV-level) energy loss with the heater on, confirming that the laser heater is responsible for seeding a high current spike in the beam which triggers the ionization of He gas on successive shots. This highlights an important feature of the mechanism for advanced accelerator applications: enabling, for example, the selective generation of current spikes in a subset of bunches in a bunch train while leaving other bunches unchanged. Such manipulations are relevant \textit{e.g.} for high repetition rate pump probe studies of plasma dynamics \cite{DArcy2022,Pompili2024}, critical for future high average power, plasma-based light sources and collider designs \cite{Galletti2024,Adli2013}.

The two datasets displayed in Fig. \ref{fig:Fig5} (c)-(d) demonstrate how the amplitude and location of the current spike in the beam can be controlled by adjusting the laser heater energy and delay. The laser energy scan presented in Fig. \ref{fig:Fig5} (c), taken with an asymmetric laser heater pulse shows a correlation between laser heater energy and two signals: the gamma ray yield from betatron radiation emitted by the beam as it undergoes transverse oscillations in the plasma channel \cite{Kiselev2004} and the emitted plasma afterglow. Both of these signals are indicative of the total amount of beam-induced ionization of He gas and the total energy deposited by the beam in the He plasma. We observe both signals rapidly increasing with laser heater energy, up to a maximum obtained at 0.18 mJ. Further increasing the pulse energy leads to a rapid suppression of both signals, indicating a decrease in the amplitude of the current spike and the amount of He ionization. 

Fig. \ref{fig:Fig5} (d) illustrates that the location of the spike in the beam can be controlled by varying the laser heater delay. The figure displays the integrated CSR signal measured during two notch scans taken with a short Gaussian laser pulse configuration at two different laser heater delays. The two datasets show two distinct regions in the notch scans where maximum suppression of CSR occurs, demonstrating that the current spike has moved to a different location in the beam. These two degrees of freedom (laser heater energy and delay), as well as laser pulse temporal profile (asymmetric, short Gaussian, etc.), enable the selective optimization of the amplitude and temporal location of the current spike in the electron bunch and further open the door to tailored current distributions with ultra-high peak current for different applications.

In conclusion, we have demonstrated the controlled generation of relativistic electron beams with ultra-high peak current and PW peak beam power using a laser-based technique to tailor the current profile at the femtosecond time-scale.  Femtosecond-level control and shaping of ultra-high current beams is a powerful tool for optimizing next generation beam interactions relevant across scientific disciplines and is poised to enable broad exploration and optimization of advanced accelerator applications. We have demonstrated that these beams have isolated current spikes with peak currents of $\sim$ 0.1 MA by measuring the coherent emission spectrum of the bunches in the UV-Visible range and by measuring the beam-induced ionization of He gas in a meter-scale plasma target. The extreme beams generated in this work have already found applications in the FACET-II experimental program for studying plasma-based injection schemes and the transition between the nonlinear and wakeless regimes in beam-driven plasma wakefields \cite{zhang2024AAC,corde2024AAC}. Future improvements may include upgrades to the spectral shaping technique to generate custom current profiles for advanced accelerator applications. Particularly interesting temporal distributions include ramped current profiles for maximizing transformer ratio \cite{lemery2015,roussel2020}, spike trains for resonant plasma excitation \cite{manwani2021} (generated \textit{e.g.} via chirped pulse beating \cite{Weling1996}) and two-bunch distributions with controllable temporal spacing for probing and driving high efficiency PWFA interactions.  
\newpage

\begin{acknowledgments}
The authors would like to acknowledge the SLAC accelerator operations and the FACET-II beam physics and operations groups for their invaluable support during the experiment. The authors also acknowledge T. Dalichaouch for his support with particle-in-cell simulations. This work was supported by the U.S. Department of Energy under DOE Contract DE-AC02-76SF00515. C. E. and K. S. also acknowledge support from the Department Of Energy Early Career Research Program.
\end{acknowledgments}

\bibliographystyle{apsrev4-1}
\bibliography{refs}

\end{document}